\documentclass[12pt,preprint]{aastex}
\newcommand{\eg}{e.g.,\,}
\newcommand{\ie}{i.e.,\,}
\newcommand{\be}{\begin{equation}}
\newcommand{\ee}{\end{equation}}
\newcommand{\bea}{\begin{eqnarray}}
\newcommand{\eea}{\end{eqnarray}}

\newcommand{\si}{\sin\phi_b}
\newcommand{\sit}{\sin\tilde{\phi}_1}
\newcommand{\sisq}{\sin^2\phi_b}
\newcommand{\sitsq}{\sin^2\tilde{\phi}_1}
\newcommand{\sicu}{\sin^3\phi_b}
\newcommand{\sitcu}{\sin^3\tilde{\phi}_1}
\newcommand{\siqu}{\sin^4\phi_b}
\newcommand{\sitqu}{\sin^4\tilde{\phi}_1}

\newcommand{\sitquin}{\sin^5\tilde{\phi}_1}
\newcommand{\sisex}{\sin^6\phi_b}

\newcommand{\cst}{\csc\tilde{\phi}_1}
\newcommand{\cssq}{\csc^2\phi_b}
\newcommand{\cstsq}{\csc^2\tilde{\phi}_1}
\newcommand{\cstcu}{\csc^3\tilde{\phi}_1}
\newcommand{\cscu}{\csc^3\phi_b}

\newcommand{\cssex}{\csc^6\phi_b}
\newcommand{\co}{\cos\phi_b}
\newcommand{\cto}{\cos\tilde{\phi}_1}

\newcommand{\ctosq}{\cos^2\tilde{\phi}_1}

\newcommand{\ct}{\cot\phi_b}
\newcommand{\ctt}{\cot\tilde{\phi}_1}
\newcommand{\mr}{\left(\frac{r_s}{2r_0}\right)}
\newcommand{\mrsq}{\left(\frac{r_s}{2r_0}\right)^2}
\newcommand{\Lr}{\Lambda r_0^2}

\begin{document}
\title{Gravitational Lensing Corrections in Flat $\Lambda$CDM Cosmology}
\author{
Ronald Kantowski\altaffilmark{1}, Bin Chen\altaffilmark{1},
Xinyu Dai\altaffilmark{1,2}
}
\altaffiltext{1}{Homer L. Dodge Department of Physics and Astronomy,
University of Oklahoma,\\
 Norman, OK 73019, USA, kantowski@nhn.ou.edu, Bin.Chen-1@ou.edu, dai@nhn.ou.edu}
\altaffiltext{2}{Department of Astronomy, University of Michigan, Ann
Arbor, MI 48109, USA}

\begin{abstract}
We compute the deflection angle to order $(m/r_0)^2$ and $m/r_0\times \Lambda r_0^2$ for a light ray traveling in a flat $\Lambda$CDM
cosmology which encounters a completely condensed mass region.
 We use a Swiss cheese model for the inhomogeneities and find that the most significant correction to the Einstein angle
occurs not because of the non-linear terms but instead occurs because the condensed mass is embedded in a  background cosmology.
The Swiss cheese model predicts a decrease in the deflection angle of
$\sim2\%$ for weakly lensed galaxies behind the rich cluster A1689,
and that the reduction can be as large as $\sim5\%$ for similar rich clusters at $z\approx 1$.
Weak lensing deflection angles caused by galaxies can likewise be reduced by as much as $\sim$4\%.
We show that the lowest order correction in which $ \Lambda$ appears is proportional to $m/r_0\times \sqrt{\Lambda r_0^2}$ and could
cause as much as a  $\sim$0.02\% increase in the deflection angle for light that passes
through a rich cluster. The lowest order non-linear correction in the mass
 is proportional to $m/r_0\times \sqrt{m/r_0}$ and can
increase the deflection angle by $\sim0.005\%$ for weak lensing by galaxies.
\end{abstract}


\section{Introduction }


Recently \citet{Rindler07} have stirred interest in the possibility of
measuring the cosmological constant $\Lambda$ through its effect on the deflection of
light that traverses large galaxy clusters by asserting that $\Lambda$ has a non-negligible effect
on small angle bending.  Several papers have since appeared to support
the existence of an effect  \citep{Ishak08a,Ishak08b,Ishak10,Sereno08,Sereno09,Schucker09a} although qualitatively disagreeing on its value and/or interpretation,
but others \citep{Park08, Khriplovich08}
contest the existence of an effect arguing that the additional bending caused by $\Lambda$ vanishes when
measured by observers moving with the Hubble flow.
We purport to give the definitive answer to this question as well as several other related ones.
When comparing observations with and without
a $\Lambda$ one must compare observations of two different sets of events,
by two different observers, in two different universes. One ideally attempts to
make common as many kinematic and dynamic properties as possible in the
two gedanken experiments. To conclude whether $\Lambda$ does or doesn't cause bending can
easily depend on what is held in common and what property is compared in the two experiments.
For example, a photon orbit in a \citet{Kottler18} spacetime (Schwarzschild with a cosmological constant)
does not depend on $\Lambda$ if static coordinates are used, see Eqs.\,(\ref{Kottler}) and (\ref{RR0}).
One could hence conclude that $\Lambda$ does not effect bending.
However, as \citet{Rindler07} point out, observers, stationary  relative to the Schwarzschild
mass, will measure an angle between the photon direction and the radial  direction
that does depends on $\Lambda$. From this observation, one could conclude that $\Lambda$ does effect bending.
Both conclusions are valid, but neither answers the outstanding question, ``How does
$\Lambda$ contribute to deflections caused by large inhomogeneities in the otherwise
homogeneous background cosmology?"
This is just one of the questions we definitively answer, subject to the condition
that the inhomogeneity is significantly condensed  and has no peculiar motion.

To correctly analyze $\Lambda$'s non-linear effect on bending we found it necessary to use exact solutions to Einstein's equations. These solutions reveal a somewhat surprising value
for the lowest order non-linear correction in the deflecting mass $m$ to the familiar Einstein deflection
formula $4Gm/c^2r_0$ [see the square root term in Eq.\,(\ref{total})].
 This correction, like the $\Lambda$ correction,
increases the deflection and occurs because  the deflector  is
embedded in a universe which expands.
By using an exact inhomogeneous cosmology the  largest correction to $4Gm/c^2r_0$ is revealed not to be a non-linear term but instead
is caused by the limited time the
deflector has to influence a passing photon [see the $\cos^3\tilde{\phi}_1$ term in Eqs.\,(\ref{total}) and (\ref{cos3})].
 The limited time or equivalently the limited range of the inhomogeneity can be thought of as a result of shielding by the homogeneous background in which the deflecting mass is embedded and decreases the deflection (relative to the Einstein value). General relativity (GR) requires that the two gravity fields, the homogeneous background and the local inhomogeneity, be appropriately matched at their bounding surface. Because the Swiss cheese models are the only known exact GR solutions that embed  spherical inhomogeneities in  expanding homogeneous universes, we use them.
Gravitational lensing calculations in cosmology are usually done by
superimposing a deflecting mass on top of a homogeneous mass density and ignoring any boundary matching.
The resulting deflection angle is obtained by a simple summation of the Einstein
expression $4Gm/c^2r_0$. The results can at best
be accurate to first order in the ratio of mass $m$ to minimum impact distance $r_0$, and if the shielding predicted by boundary matching in Swiss cheese is accurate, the linear term can be in error by a few percent in physically realizable circumstances (see Figure 4).

Because our goal is to correctly present the higher order corrections  and because
a simple superposition of the masses isn't satisfactory, we resort to using exact
solutions to Einstein's gravity (see Section 2).
It is in the non-linear corrections that the cosmological constant first appears. What we calculate in this paper
is (in a series approximation) the angle $\alpha\equiv\xi_2-\xi_1$ (see Fig. 1) between two spatial directions $\xi_2$ and $\xi_1$ as seen by co-moving observers in a flat Robertson-Walker (RW) spacetime, where $\xi_1$ is the spatial direction of a photon before it encounters an inhomogeneity
here described by the vacuum Kottler spacetime [see Eq.\,(\ref{Kottler})] and $\xi_2$ is the direction of the photon after it has emerged from the Kottler condensation. The dynamics of the RW metric is determined by General Relativity sourced by pressure free matter (often referred to as dust) and includes a cosmological
constant. These are relatively simple Friedman-Lema\^itre-Robertson-Walker (FLRW)
 cosmological models, see Eq.\,(\ref{RW}).
Because the RW cosmology used is spatially flat and non-rotating  (a) the angle between any two co-moving
spatial directions is well defined independently of when or where the directions are measured, and (b)
the spatial direction of an undeflected photon remains fixed.

In Section 2 we describe the inhomogeneous cosmology we use and in Section 3 we outline some
details of how we compute the bending angle of a photon that encounters an inhomogeneity.
In Section 4 we discuss limits on the usefulness of our results and
compare the Einstein angle with our corrected results
for deflections caused by inhomogeneities ranging from
galaxies to rich clusters.

\section{Swiss Cheese: Locally Inhomogeneous Cosmologies}


We  use a single condensation in a Swiss cheese cosmology to compute light deflections caused by
local inhomogeneities so there can be no doubt about errors
introduced by gravity approximations.
Because Swiss cheese is an exact solution to Einstein's
equations \citep{Einstein45,Schucking54} its use insures the accuracy of the superposed gravity field
and automatically takes into account the finite range of the mass density perturbation as well
as observer aberration. The model was first used by \citet{Kantowski69a} without the cosmological constant
to settle the dispute as to whether or not inhomogeneities effected mean luminosities.
At that time, the validity of predictions \citep{Zeldovich64,Dashevskii65,Dashevskii66,Bertotti66,Gunn67}
obtained using approximate GR solutions, which indicated that mass
concentrations caused the average distance-redshift  relation to differ
from the pure homogeneous value, were being questioned. Even though the results
are on occasion doubted by some, the Swiss cheese model gave the definitive answer, there is an effect, see \citet{Kantowski95}.
This model again comes to the rescue by clearly demonstrating the extent to which
 the cosmological constant $\Lambda$ influences the small angle bending of a photon that passes a single
mass concentration, see Eq.\,(\ref{total}). Even though others have computed bending angles that depended
on  $\Lambda$, until now, questions abound as to their  usefulness and/or accuracy in a cosmological setting.
We have succeeded in  giving a rigorously derived expression for
this deflection.

The Swiss cheese model simultaneously accounts for the finite size of the
deflector's influence, the motion of
the cosmic observers, and the non-linear effects of gravity.
The basic idea of Swiss cheese is to
remove non-overlapping co-moving spheres of homogeneous dust
from one of the homogeneous FLRW cosmologies and replace them with  gravity
fields representing appropriately condensed spherical mass distributions. If the cosmology
is without a cosmological constant the simplest replacements are Schwarzschild metrics
and if there is a $\Lambda$ the simplest replacements are Kottler metrics \citep{Dyer74}. These
condensations are the most extreme for Swiss cheese.
An infinity of less extreme models can be constructed by using the
Lema\^{\i}tre-Tolman-Bondi metrics \citep{Lemaitre33,Tolman34,Bondi47} to represent spherically symmetric
dust concentrations \citep{Kantowski69b}.
In all Swiss cheese models
the metric that is used to fill a dust condensation must match first and second
fundamental forms on the boundary. In the case of Schwarzschild the metric's mass is
fixed by the dust's density and the size of the condensed hole, and in the case of Kottler, $\Lambda$
is additionally required to be the same inside as out.
In this calculation we stick with the extreme but unique condensation, the
Kottler metric, to arrive at a unique deflection angle. \citet{Schucker09b} uses this same model
but because he only considers a single numerical example his results are difficult to compare with ours.

The two metrics are: outside, flat ($\Omega_\Lambda+\Omega_m=1$) FLRW
\be
ds^2=-c^2dT^2+R(T)^2[d\chi^2+\chi^2(d\theta^2+\sin^2\theta\, d\phi^2)],
\label{RW}
\ee
with the cosmic time development given by the Friedman equation
\be
\frac{\dot{R}}{R}=H_0\sqrt{\Omega_\Lambda+\Omega_m\left(\frac{R_0}{R}\right)^3},
\ee
and inside, the static Kottler metric \citep{Kottler18}
\be
ds^2=-\gamma(r)^{-2}c^2dt^2+
\gamma(r)^2dr^2+
r^2(d\theta^2+\sin^2\theta\, d\phi^2),
\label{Kottler}
\ee
where $\gamma(r)$ is defined by
\be
\gamma(r)\equiv 1/\sqrt{1-\frac{r_s}{r}-\frac{\Lambda r^2}{3}}.
\label{gamma}
\ee
Boundary matching at co-moving FLRW radius $\chi_b$ constrains the Schwarzschild radius $r_s$
of the condensed mass to be
\be\label{match1}
r_s=\Omega_m\frac{H_0^2}{c^2}(R_0\chi_b)^3 ,
\ee
and the additional Kottler parameter $\Lambda$ to coincide with the FLRW value, \ie
\be
\Lambda=3\Omega_\Lambda\frac{H_0^2}{c^2}.
\ee
The Kottler and RW angular coordinates are matched at the boundary and  the
radius of the Kottler hole expands according to
\be
r_b(T)=R(T)\chi_b
\label{rb}
\ee
 (for some numerical examples see the Mass and $r_b$ columns of Table 1).
As seen by a stationary Kottler observer the dust boundary of the Kottler hole moves
with Lorentz parameters $\beta_b$ and $\gamma_b$ given by
\bea
\label{Lorentz}
\gamma_b&\equiv&\gamma(r_b),\\
\beta_b&\equiv&\sqrt{1-\gamma_b^{-2}}=H_br_b/c,\nonumber
\label{beta}
\eea
where $H_b$ is the time dependent Hubble parameter of the boundary.
The normalized 4-velocity of the boundary coincides with the RW co-moving dust velocity $\hat{u}_{RW}$
at the boundary
and is of the form
\be
\hat{u}_{RW}=\gamma_b\,\hat{u}_K+\beta_b\gamma_b\,\hat{r}_K,
\label{boundary}
\ee
when expressed in terms of unit Kottler time and radial vectors, respectively $\hat{u}_K$ and  $\hat{r}_K.$

\section{The Photon's Path}

In Fig.\,1  we show the spatial orbit of a slightly deflected photon $r(\phi)$ that enters and exits a Kottler condensation. The coordinates have been rotated
to put the orbit in the $\theta=\pi/2$
plane and to make it symmetric  about $\phi=\pi/2$ while in Kottler.
The tangent to the photon's geodesic path is
\be
k=\frac{\ell}{r_0}\left[\frac{\gamma(r)}{\gamma_0}\hat{u}_K
\pm\sqrt{\frac{\gamma(r)^2}{\gamma_0^2}-\frac{r_0^2}{r^2}}\ \hat{r}_K-\frac{r_0}{r}\hat{\phi}_K\right],
\label{photon-Kottler}
\ee
where $\ell$ is an angular momentum like constant; $r_0$ is the minimum $r$
and occurs at $\phi=\pi/2$; $\gamma_0\equiv\gamma(r_0)$; and $\hat{u}_K,\ \hat{r}_K $
and  $\hat{\phi}_K$ are unit vectors pointing respectively in the static time, radial and azimuthal Kottler directions.
The actual orbit $r(\phi)$ is approximated as
\bea
r/r_0&=&\csc\phi\Biggl\{1-\left(\frac{r_s}{2r_0}\right)\bigl[-1+2\csc\phi-\sin\phi\bigr]\nonumber\\
&&+\left(\frac{r_s}{2r_0}\right)^2
\left[-\frac{17}{4}+\frac{15}{4}(\phi-\frac{\pi}{2})\cot\phi+4\csc^2\phi+
\frac{1}{4}\sin^2\phi\right]\Biggr\}+{\cal O}\left[\left(\frac{r_s}{2r_0}\right)^3\right] .
\label{RR0}
\eea
For this to be a valid expansion not only must $r_s/r_0<<1$ but $\phi$ must also satisfy $\sin\phi>>r_s/r_0$.
The tangent to  the photon as it travels in the $\theta=\pi/2$ plane of a flat RW spacetime is of the form
\be
k=\frac{{\rm con}}{R(T)}\left[\hat u_{RW}+\cos(\phi-\xi)\hat\chi-\sin(\phi-\xi)\hat\phi_{RW}\right],
\ee
where $\hat u_{RW},\hat\chi,$ and $\hat\phi_{RW}$ are respectively unit co-moving time,
radial, and azimuthal vectors in RW. The significance of the constant angle $\xi$ is that $\tan\xi$
is the slope of the photon's straight line orbit in the co-moving x-y plane, see Fig.\,1.
When the photon's tangent vector is matched across the boundary of the dust hole the following single
(exact) constraint results:
\be
\Bigl[1+\beta_b\,\cos(\phi_b-\xi)\Bigr]\frac{r_0}{r_b}=\frac{\sin(\phi_b-\xi)}{\gamma_0},
\label{photon-match}
\ee
where $r_b,\phi_b,$ and $\beta_b$ are evaluated at the photon's entrance/exit point
on  the boundary of the Kottler hole.
From Eq.\,(\ref{photon-match}) we obtain the following exact expression for  $\xi$
\be
\sin\xi=
\left\{
-B\ \frac{r_0}{r_b}
\pm A
\sqrt{
1-\left(\frac{r_0}{r_b}\right)^2-\frac{r_s}{r_0}\left[1-\left(\frac{r_0}{r_b}\right)^3\right]
}
\right\}
/\left\{1-\frac{r_s}{r_0}\left[1-\left(\frac{r_0}{r_b}\right)^3\right]\right\},
\label{xi}
\ee
where
\bea
A&\equiv&\co\,\beta_b\,\frac{r_0}{r_b} -\frac{\si}{\gamma_0},\nonumber\\
B&\equiv&\si\,\beta_b\,\frac{r_0}{r_b} +\frac{\co}{\gamma_0}.
\eea
The $-A$ choice is made in Eq.\,(\ref{xi}) at the exit point and the Kottler coordinates on the boundary are
taken as $r_b=r_2,\phi_b=\phi_2$. The $+A$ choice is made at the entrance point and the Kottler coordinates are
taken as $r_b=r_1,\phi_b=\pi-\tilde{\phi}_1$, ($\tilde{\phi}_1$ is the supplement of the entrance
azimuthal coordinate, see Fig.\,1).

In what follows we give some of the details necessary to approximately evaluate the
deflection angle $\alpha=\xi_2-\xi_1$ caused by encountering a condensation in the homogeneous dust.
The reader not interested in the details  can jump to the result
called  $\alpha_{\rm total}$ given in Eq.\,(\ref{total}). The calculation is somewhat complicated because the
Kottler hole expands as it is traversed by the photon.  The deflection angle naturally divides into a
part that depends on the initial size of the hole [$\alpha_{\rm static}$ given in Eq.\,(\ref{static})]
and an additional part
 caused by the extended path
required of the photon  to exit the expanded hole [$\alpha_{\rm expand}$ given in Eq.\,(\ref{expand})].
The extended path, described by  $\Delta r$ and $\Delta\phi$,
is  given in Eqs.\,(\ref{Delta-r}) and (\ref{Delta-phi}), see Fig.\,1.
We will see that the expansion part
gives the most significant nonlinear part of the correction to the familiar Einstein term $4Gm/c^2r_0$.

\subsection{Approximation Details}

To compute the photon's direction $\xi$ in the dust approximately we assume both
$\Lr$ and $r_s/r_0$ are small (perhaps even of the same order) and expand Eq.\,(\ref{xi}) in the two small parameters
\be
\delta\equiv\sqrt{\frac{\Lr}{3}+\frac{r_s}{r_0}\left(\frac{r_0}{r_b}\right)^3}=\beta_b\frac{r_0}{r_b},
\label{delta}
\ee
and
\be
\delta_m^2\equiv\frac{r_s}{r_0}\left(\frac{r_0}{r_b}\right)^3.
\label{delta-m2}
\ee
The result is

\be
\xi=-\delta+C_2(\phi_b)\delta_m^2-\frac{1}{6}\delta^3+C_3(\phi_b)\delta_m^2\delta+
C_4(\phi_b)\delta_m^4+{\cal O}\bigl(\delta^5\bigr),
\label{xi-approx}
\ee
where the coefficients are defined by
\bea\label{C}
C_2(\phi_b)&\equiv& -\ct\left(\frac{1}{2}+\cssq\right),\\
C_3(\phi_b)&\equiv& \frac{1}{2}\left(1-\cscu\right),\nonumber\\
C_4(\phi_b)&\equiv&\cssex\left[\frac{15}{32}(2\phi_b-\pi)+\ct\left(3-\si-\frac{15}{16}\sisq- \frac{1}{2}\sicu+
\frac{1}{8}\siqu+ \frac{1}{4}\sisex\right)\right].\nonumber
\eea
From Eq.\,(\ref{xi}) we can conclude that when $r_s \rightarrow 0, \sin\xi \rightarrow -\delta = -\sqrt{\Lr/3}$
exactly with no dependence on $r_b$ or $\phi_b$. This limit is consistent with Eq.\,(\ref{xi-approx}).
The conclusion is that when $r_s =0$ there is no $\Lambda$ bending. This is an obvious conclusion
because the spacetime inside and outside of the hole is exactly the same,
\ie no physical difference inside and out exists. The only difference is  in which
coordinates are being used.

When $0<r_s<< r_0$ we proceed by eliminating $r_1$ and $r_2$ using Eq.\,(\ref{RR0}) and then
expanding $\phi_2$ about $\tilde{\phi}_1$ by writing
\be
\phi_2=\tilde{\phi}_1+\Delta\phi.
\ee
This gives us two terms to evaluate
\be
\alpha_{\rm static}\equiv \xi_2(\tilde{\phi_1})-\xi_1(\pi-\tilde{\phi_1}),
\ee
and
\be
\alpha_{\rm expand}\equiv\left(\frac{d\xi}{d\phi}\right)_{\tilde{\phi}_1}\Delta\phi+\frac{1}{2}\,\left(\frac{d^2\xi}{d\phi^2}\right)_{\tilde{\phi}_1}(\Delta\phi)^2+\frac{1}{6}\,\left(\frac{d^3\xi}{d\phi^3}\right)_{\tilde{\phi}_1}(\Delta\phi)^3+
{\cal O}\left(\Delta\phi\right)^4.
\label{alpha-expand}
\ee
The first term can be evaluated immediately using Eqs.\,(\ref{xi-approx}) and (\ref{C})
giving  the $\Lambda$ independent expression
\bea
\alpha_{\rm static}&=&
-2 \mr\cto \left[2+\sitsq\right]+
\mrsq
\Biggl[
\frac{15}{4}(2\tilde{\phi}_1-\pi)+\nonumber\\
&&\cto \left(4  -\frac{15}{2}\sit+2\sitsq +7 \sitcu+2 \sitquin \right)
\Biggr]+{\cal O} \mr^3.
\label{static}
\eea
By overlooking the expansion term one would obviously conclude that there is no $\Lambda$ bending.
To evaluate $\alpha_{\rm expand}$, the expansion's contribution to bending, requires that we compute $\Delta\phi$ (or equivalently $\Delta r$)
caused by the expansion of the Kottler hole
that took place during the time it took the photon to transit the hole.
In Fig. 2 we indicate how we compute $\Delta r$. We start by giving the entrance radius $r_1$
and look for the common solution to the boundary expansion $r_b(t)$ and the photon's radial coordinate $r_p(t)$,
\ie we put
\be
c\int dt=c\int_{r_1}^{r_2} \left(\frac{dr_b}{dt}\right)^{-1}dr=
c\int_{r_1}^{r_2}\left(\frac{dr_p}{dt}\right)^{-1}dr.
\ee
We rewrite the time it takes the photon to cross the hole as the sum of the time it takes to cross from $r_1$
on the left to $r_1$ on the right  plus the extra time it takes to go from $r_1$ on the right to $r_2=r_1+\Delta r$.
We then move this last time difference to the left hand side and obtain
the following equation to solve
\be
c\int_{r_1}^{r_2} \left[\left(\frac{dr_b}{dt}\right)^{-1}-\left(\frac{dr_{p}}{dt}\right)^{-1}\right]dr
=2c\int_{r_0}^{r_1}\left(\frac{dr_p}{dt}\right)^{-1}dr=2c\int_{\pi/2}^{\tilde{\phi_1}}\left(\frac{d\phi_p}{dt}\right)^{-1}d\phi.
\label{LHS=RHS}
\ee
The right hand side $RHS$ is evaluated approximately using Eqs.\,(\ref{photon-Kottler}) and (\ref{RR0}) to obtain
\bea
RHS&=&2r_0\Biggl\{
\ctt+
\mr\left[
\ctt\left(1-2\cst\right)-2\log\left(\tan\frac{\tilde{\phi_1}}{2}\right)
\right]+\nonumber\\
&&\frac{\Lr}{18}\ctt\left[1+2\cstsq\right]
+{\cal O}\left[\mr+\Lr\right]^2
\Biggr\}.
\label{RHS}
\eea
We call the two terms on the left hand side of Eq.\,(\ref{LHS=RHS}) $LHS_b$ and $LHS_p$
and evaluate $LHS_b$ by expanding in $\Delta r$
\be
LHS_b=\left(\frac{\gamma^2_b}{\beta_b}\right)_{r_1}\,(\Delta r)+
\frac{1}{2}\frac{d\ }{dr_b}\left(\frac{\gamma^2_b}{\beta_b}\right)_{r_1}(\Delta r)^2
+
\frac{1}{6}\frac{d^2\ }{dr_b^2}\left(\frac{\gamma^2_b}{\beta_b}\right)_{r_1}(\Delta r)^3+
{\cal O}(\Delta r)^4,
\label{LHSb}
\ee
where $\beta_b$ and $\gamma_b$ as functions of $r_b$ are defined in Eq.\,(\ref{Lorentz}).
Equation (\ref{RR0}) can be used to convert $\Delta r$ into $\Delta\phi$ resulting in
\bea
 \Delta r&=&r_0\Biggl\{
-\cto\cstsq\left[1+\mr\left(1-4\cst\right)+{\cal O}\mr^2\right]\Delta\phi+\nonumber\\
&&\frac{1}{2}\Biggl[\cstcu\left(2-\sitsq\right)+{\cal O}\mr\Biggr](\Delta\phi)^2
+\nonumber\\& &
+\frac{1}{6}\Biggl[\ctt\cst(1-6\cstsq)+{\cal O}\mr\Biggr](\Delta\phi)^3+{\cal O}(\Delta\phi)^4\Biggr\}.
\label{Delta-r}
\eea
The second term on the left hand side of Eq.\,(\ref{LHS=RHS}) can be evaluated by using $\phi_p(t)$
from Eq.\,(\ref{photon-Kottler}) rather than $r_p(t)$ (just as was done with $RHS$) and gives
\bea
LHS_p=r_0\cstsq\Biggl\{
\Biggl[1+{\cal O}\left(\frac{r_s}{r_0}+\Lr\right)\Biggr](\Delta\phi)-\ctt\Biggl[1+{\cal O}\left(\frac{r_s}{r_0}+\Lr\right)\Biggr](\Delta\phi)^2+{\cal O}\left(\Delta\phi\right)^3
\Biggr\}.
\label{LHSp}
\eea

\subsection{The Resulting Deflection}

Combining Eqs.\,(\ref{RHS}), (\ref{LHSb}) and (\ref{LHSp}) in Eq.\,(\ref{LHS=RHS})
we obtain the change that occurs in the exiting value of $\phi$, \ie \
$\Delta\phi\equiv\phi_2-\tilde{\phi}_1$, caused by the expansion of the hole's boundary as the photon traverses
\bea
\Delta\phi &=&-2\beta_1 \sit +\left(\frac{r_s}{r_0}\right)\Biggl[3\cto\sitsq\nonumber\\
&&-
\beta_1\left(2+\frac{7}{3}\sitsq-6\sitqu
-2\log\left\{\tan\frac{\tilde{\phi}_1}{2}\right\}\tan\tilde{\phi}_1\sit\right)\Biggr]+\nonumber\\
&&-
\frac{1}{9}\,\beta_1\,\Lr\sit+
{\cal O}\left(\frac{r_s}{r_0}+\Lr\right)^2,
\label{Delta-phi}
\eea
where $\beta_1$ is the expansion velocity ($v/c$) of the dust as seen by observers (who are stationary
relative to the condensed mass) at the time the photon enters the Kottler hole, see Fig. 1.
Inserting this into Eq.\,(\ref{alpha-expand}) we have the additional deflection angle $\alpha_{\rm expand}$ caused by the extended trajectory
of the photon in the Kottler void
\bea\label{expand}
\alpha_{\rm expand}&=&\mr \cto
\left[6 \sitsq
-12\, \cto\sit \sqrt{\frac{\Lr}{3}+\frac{r_s}{r_0}\,\sitcu} +\Lr \left(\frac{8}{3}-\frac{20}{3} \sitsq \right)
\right]\cr
& &+\mrsq\Biggl[6\cto \left(4 \sit - \sitsq +2 \sitcu-11 \sitquin \right)-12 \log\left\{\tan\frac{\tilde{\phi}_1 }{2}\right\} \sitcu\Biggr]\cr
& &+
{\cal O}\left(\frac{r_s}{r_0}+\Lr\right)^{5/2}.
\eea
Combining Eqs.\,(\ref{static}) and (\ref{expand}) we obtain the total bending angle $\alpha_{\rm total}$ caused by a photon
entering and exiting a Kottler condensation
\bea\label{total}
\alpha_{\rm total}&=&\mr \cto
\left[-4 \ctosq
-12\,\cto\sit\sqrt{\frac{\Lr}{3}+\frac{r_s}{r_0}\,\sitcu} +\Lr \left(\frac{8}{3}-\frac{20}{3} \sitsq \right)
\right]\nonumber\\
& &+\mrsq\Biggl[ \frac{15}{4}(2\tilde{\phi}_1-\pi)
+\cto \left(4+\frac{33}{2} \sit -4 \sitsq +19 \sitcu-64 \sitquin \right)\nonumber\\
&&-12 \log\left\{\tan\frac{\tilde{\phi}_1 }{2}\right\} \sitcu
\Biggr]+
{\cal O}\left(\frac{r_s}{r_0}+\Lr\right)^{5/2}.
\eea
The reader should observe that a negative contribution to the bending angle
is towards the lens and a positive is away.
Also recall that these approximate expressions were derived assuming $\sit>>r_s/r_0$.
Our deflection angle accounts for the finite time (equivalently range) that gravity has to act on the passing photon
as well as aberration effects caused by switching between moving observers.
A finite range is equivalent to a shielding of
the perturbation's mass by the homogeneous distribution of its neighbors, \ie beyond $r_b$ of Eq.\,(\ref{rb})
the effect of the neighbors completely suppress effects of the inhomogeneity.
The deflection angle $\alpha_{\rm total}$ appropriately vanishes in the limit $\tilde{\phi}_1\rightarrow \pi /2$, \ie when the photon only grazes the condensation, and for small $\tilde{\phi}_1$ the lowest order term in the bending angle approaches the Einstein value $4Gm/c^2r_0$ as expected.  For an arbitrary impact $\tilde{\phi}_1$, however, the linear term in $\alpha_{\rm total}$ is
\be\label{cos3}
\alpha_{\rm linear}=-4\mr\cos^3\tilde{\phi}_1,
\ee
and, in some weak lensing circumstances,  predicts potentially detectable differences from the Einstein value.
 In standard lensing calculations  [see \cite{Bourassa75}] there is no attempt to make the deflector's gravity field part
of the cosmology's gravity field as GR really requires. Deflector masses are simply taken as additions to the cosmology's mean mass density
and consequently have `$\infty$' range.
Swiss cheese,
the only known and relevant GR solution, makes the deflector mass a contributor to the cosmology's mean density and as
a consequence, the gravity field of the deflector is limited in range. This limited range is seen to be
important when the the impact angle is above a tenth of a radian (Figure 4).

Another somewhat surprising result is that the lowest order correction to the Einstein value,
other than the finite time effect represented by the dependence on $\tilde{\phi}_1$, is
the dependence on the expansion rate, \ie the square root term in Eqs.\,(\ref{expand}) and
(\ref{total}) [see Eq.\,(\ref{delta})]. We can interpret the source of this term as the extra
time (or equivalently distance) the Schwarzschild mass has to act on the passing photon.
The Kottler hole expands
in size as the photon traverses, and since the
cosmological constant contributes to the Hubble expansion it contributes to the extra time.
Others have also argued that $\Lambda$ effects $\alpha$, \eg \citet{Sereno09} finds a $\Lambda$
contribution to small angle bending of order $(r_s/r_0)\Lambda r_0^2$ which we do find even if of
opposite sign and differing amount, \citet{Ishak08a,Ishak08b}
find a term of order  $\Lambda r_0r_b\sim \Lambda r_0^2\cst$ which we do not.
The most important $\Lambda$ correction we find, \ie the square root term in Eq.\,(\ref{total}),
seems to have gone undetected by others because of
the approximations they used.
In the next section we estimate just how important these
corrections to the Einstein result can be.

\section{Discussion}

In Fig.\,3 we have plotted three sets of bending angles for three deflecting masses
ranging from a large galaxy mass to a rich cluster mass, respectively
$10^{11}M_\sun$\,(lower in red), $10^{13}M_\sun$\,(middle in green),
and  $10^{15}M_\sun$\,(top in blue) all at redshift $z=1$.
Note that redshift $z$ plays a part because redshift influences the
entrance/exit size of the Kottler hole, see Eqs.\,(\ref{match1}) and (\ref{rb}).
The cosmological parameters we used are $\Omega_m=0.3,$ $\Omega_\Lambda=0.7,$ and $H_0=70\,{\rm km}\,{\rm s}^{-1}{\rm Mpc}^{-1}.$  For each mass we have plotted four bending angles in arcseconds as functions of
$\tilde\phi_1$ (the supplement of the azimuthal impact angle).  The thick lines are $|\alpha_{\rm total}|$ of Eq.\,(\ref{total}),
the short dashed lines are $|\alpha_{\rm static}|$ of Eq.\,(\ref{static}),
the dashed lines are $\alpha_{\rm expand}$ of Eq.\,(\ref{expand}),
and the  thin solid lines are the Einstein values $2r_s/r_0$.
All deflection angles are negative (attractive), \ie towards the deflector, except
$\alpha_{\rm expand}$ which is away from the deflector. Because of the log-log scale it was necessary to plot
absolute magnitudes \ie
 $|\alpha_{\rm total}|=|\alpha_{\rm static}|-\alpha_{\rm expand}$. The reader can easily see (to the accuracy of the
plot) that if
$\alpha_{\rm expand}$ is neglected the deflection angle follows $2r_s/r_0$ out to  $\sim 30^\circ$, however, when
$\alpha_{\rm expand}$ is included the deflection angle follows $2r_s/r_0$ only out to  $\sim 10^\circ$.
This observation is clearly independent of the masses shown and in fact is quite
independent of the deflectors' redshifts.
The fractional difference in  $\alpha_{\rm total}$ and the Einstein value plotted in Fig.\,4 is independent
of the deflector's mass  ($10^{11}-10^{15}M_{\sun}$) and redshift ($0<z<2$) for the range of $\tilde\phi_1$ plotted.
Noticeable redshift dependent differences would begin to appear for the three masses only below $\tilde\phi_1\sim 2^\circ$.
From Fig.\,4 we can conclude that for angles above $\sim 4^\circ$ the fractional differences of $\alpha_{\rm expand}$
and the Einstein values are greater than 1\% and above $\sim 40^\circ$ the differences are above 100\%.

In Table~\ref{tab:corr} we use our corrected bending angle Eq.\,(\ref{total}) to estimate corrections in  bending angles for strong and weak lensing by clusters and elliptical galaxies. We look at the following cases: the large image separation cluster lens Abell~1689 at $z=0.18$, the high redshift cluster RDCS~1252$-$2927 at $z=1.24$, and a typical $z=0.5$ elliptical galaxy. In A1689, we calculate the bending angle corrections for the largest arc separation of 45\arcsec\ for strong lensing and weak lensing measurements at 10\arcmin\ away from center (Umetsu \& Broadhurst 2008)  by using the mass profile of recent X-ray measurements (Peng et al.\ 2009).
We also calculate the correction in a high redshift cluster RDCS~1252$-$2927, where the weak lensing signals have been detected out to 3\arcmin\ (Lombardi et al.\ 2005).
For lensing by galaxies, we choose a typical elliptical galaxy at $z=0.5$ and use the mass profile and weak lensing detections in Gavazzi et al.\ (2007).
In general, we find the corrections in the bending angles for strong lensing are quite small, \eg the largest correction ($1-\cos^3{\tilde{\phi}_1}$) is just 0.07\%  for the largest separated arcs in A1689.
However, for weak lensing, the correction can reach 2\% for the weak lensing signals detected in the
outermost regions of the cluster in A1689, and the correction can reach 5\% for the $z=1.24$ cluster RDCS~1252$-$2927.
For the weak lensing signals detected using an ensemble of elliptical galaxies (Gavazzi et al. 2007), the correction is 4\% for the outermost bin.
A correction of this amount will present an additional challenge for using weak lensing as a tool for precision cosmology.
For the corrections involving the $\Lambda$ term, the largest is 0.02\% for  weak lensing in high redshift clusters which is not detectable in current observations.
We expect our model to be relevant for weak lensing induced by the large scale structure including weak lensing of the cosmic microwave background, where even larger volumes are involved.
We expect a large correction due to the $1-\cos^3{\tilde{\phi}_1}$ term and a presumably detectable correction involving $\Lambda$.

The corrections we give for strong lensing are negligible because $\tilde{\phi}_1$ is small and only a small fraction of the inhomogeneous mass appears inside the Einstein ring.
For these cases, our corrections may not be accurate because the effective lensing mass is not spherically distributed as it is in our model. Non-linear corrections are conceivably sensitive to the difference in cylindrical and spherical symmetry.
More realistic models are needed to fully constrain corrections for strong lensing.
In general, the applicability of the corrected deflection angle $\alpha_{\rm total}$ in Eq.\,(\ref{total}) is limited
to spherical inhomogeneities, the majority of whose
mass is within the minimum impact of the light ray.
This is because we used a fully condensed Swiss cheese model, \ie the homogeneity is
represented by a Schwarzschild mass. Because we are calculating non-linear corrections one cannot expect
Eq.\,(\ref{total}) to give an accurate answer by simply including that fraction of the  mass within the
impact cylinder as is normally done in lensing. Consequently more accurate mass profiles in the Swiss cheese would be appropriate for the strong lensing examples in Table 1.

Work on this paper was initiated to correctly quantify the cosmological constant's effect on small angle deflections of photons caused by mass inhomogeneities in an otherwise homogeneous cosmology. By using an exact solution to GR we
established that $\Lambda$'s effect is non-linear thus requiring
use of a gravity theory beyond Newton's. The model we used, a flat Swiss cheese cosmology,
also predicts a significant decrease in the deflection angle
caused by the shielding of an inhomogeneity by its homogeneously distributed neighbors.
Shielding occurs because the deflectors are contributors to the cosmology's mean density.
Standard lensing calculations completely overlook shielding because deflectors are treated as additions to the mean.

Perturbations to  $\alpha_{\rm total}$ would obviously exist if the neighbors generated a shear at the
site of the deflector. The accuracy of Swiss cheese predictions depends on the scale at which inhomogeneous matter follows
the background Hubble flow, \ie on what scale the cosmological principle is satisfied. The simple Swiss cheese
model used here doesn't allow for peculiar motions but does account for
the scale of the cosmology effected by an inhomogeneity, \ie beyond $r_b$ the perturbed spacetime returns to the mean cosmic flow.
In the neighborhood of the Local Group, where good observational data is available, most galaxies follow the Hubble flow with
only small deviations (e.g., Karachentsev et al.\  2009).
 A hierarchical Swiss cheese condensation could be used to include shear and peculiar motion
but it would not only complicate this calculation
by introducing several additional parameters, it would most certainly obscure the source of the
$\Lambda$ term in the results.
To keep the result as simple as possible we did not attempt to
estimate the size of these additional perturbations.

Our results \eg Eq.\,(\ref{total}) are stated in terms of the parameters $\tilde{\phi}_1$
and $r_0$ described in
Section 3 and Figure 1 and are not necessarily the most convenient ones to use in
lensing applications, however, they were convenient for the above derivations. To have the incoming
photon travel parallel to the x-axis one only has to rotate the coordinates clockwise an amount $\xi_1$ given
in Eq.\,(\ref{xi}).

\begin{deluxetable}{cccccccccc}
\tabletypesize{\scriptsize}
\tablecolumns{10}
\tablewidth{0pt}
\tablecaption{Examples of Gravitational Lensing Corrections in $\Lambda$CDM Cosmology \label{tab:corr}}
\tablehead{
\colhead{Name} &
\colhead{Lensing} &
\colhead{redshift} &
\colhead{Mass} &
\colhead{$r_b$} &
\colhead{Impact Angle} &
\colhead{$\tilde{\phi}_1$} &
\colhead{$1-\cos^3{\tilde{\phi}_1}$} &
\colhead{ratio1\tablenotemark{a}} &
\colhead{ratio2\tablenotemark{b}}
\\
\colhead{} &
\colhead{} &
\colhead{} &
\colhead{($M_{\odot}$)} &
\colhead{(Mpc)} &
\colhead{(arcsec)} &
\colhead{(degrees)} &
\colhead{} &
\colhead{} &
\colhead{}
}
\startdata
A1689 & strong & 0.18 & $8\times10^{13}$ &           6.6  & 45  & 1.2 & 0.00065 & $2.2\times10^{-6}$ & 1.4  \\
A1689 & weak   & 0.18 & $10^{15}$        &           15.3 & 600 & 6.8 & 0.021   & 0.00017 & 1.4  \\
RDCS1252$-$2927 & weak & 1.24 & $10^{15}$ &           8.0 & 180 & 11 & 0.052   & 0.00040 & 0.20  \\
Elliptical Galaxy & strong & 0.5 & $3\times10^{11}$ & 0.8 & 2 & 0.87 & 0.00035 & $1.7\times10^{-7}$ & 0.69  \\
Elliptical Galaxy & weak   & 0.5 & $10^{13}$ &        2.6 & 70 & 9.6 & 0.041 & $6.6\times10^{-5}$ & 0.69  \\
\enddata
\tablenotetext{a}{ratio1 $ \equiv 4 \tan{\tilde{\phi}_1} \sqrt{\frac{\Lambda r_0^2}{3} + \frac{r_s}{r_0} \sin^3{\tilde{\phi}_1}}$ [the ratio of the  next order correction to the lowest order term, see Eq.\,(\ref{total})].}
\tablenotetext{b}{ratio2 $ \equiv \frac{\frac{\Lambda r_0^2}{3}}{\frac{r_s}{r_0} \sin^3{\tilde{\phi}_1}}$ (measures the relative importance of $\Lambda$ in the square root term). }
\end{deluxetable}

\acknowledgements{}
This work was supported in part by NSF grant AST-0707704 and US DOE
Grant DE-FG02-07ER41517. B. Chen also thanks the University of Oklahoma Foundation for a fellowship.

\clearpage

\figcaption{}
A photon travels left to right entering a Kottler hole at $r=r_1,\phi=\pi-\tilde{\phi}_1 $
and returns  to the FLRW dust at $r=r_2,\phi=\phi_2$. The photon's orbit has been chosen
symmetric in Kottler about the
point of closest approach $r=r_0,\phi=\pi/2$. Due to the cosmological expansion $\Delta r\equiv r_2-r_1>0$.
The slope of the photon's co-moving trajectory in the x-y plane is $\xi_1$
when incoming and $\xi_2$ after exiting. The resulting deflection angle as seen by a co-moving observers
in the FLRW background is $\alpha=\xi_2-\xi_1$.
\label{fig1}

\epsscale{1.0}

\plotone{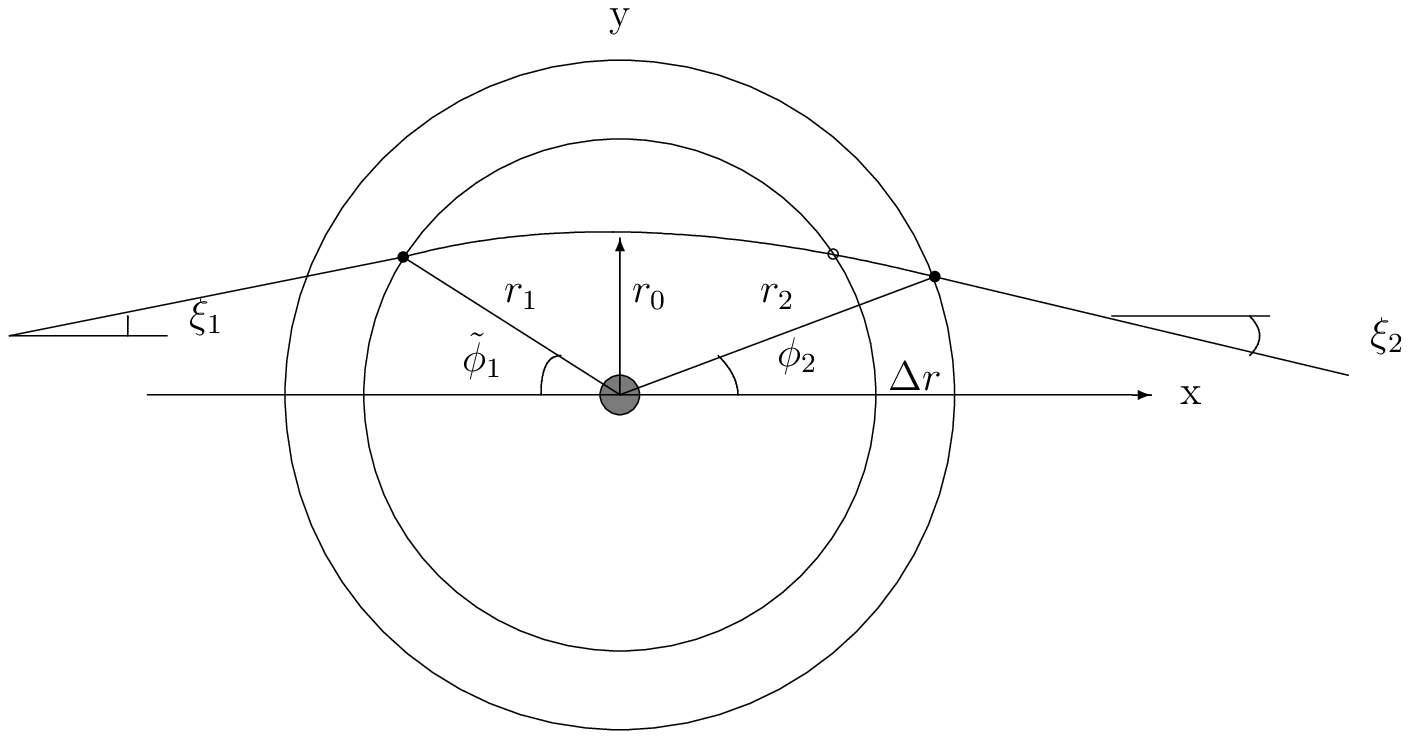}
\begin{center} Figure 1 \end{center}

\eject

\figcaption{}
While a photon travels through a Kottler hole entering at $r=r_1,\phi=\pi-\tilde{\phi}_1 $
and exiting at $r=r_2,\phi=\phi_2$ its  radial coordinate varies with time as $r_p(t)$ and
the boundary of Kottler hole continually expands according to $r_b(t)$. The exit coordinates  differ from
the entrance values by $\Delta\phi=\phi_2-\tilde{\phi}_1$ and $\Delta r= r_2-r_1$.
\label{fig2}

\epsscale{1.0}

\plotone{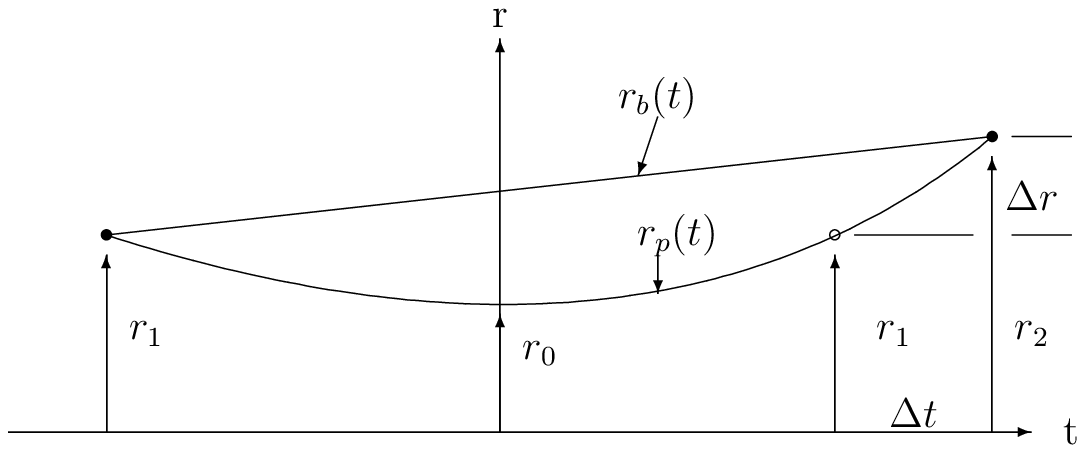}
\begin{center} Figure 2 \end{center}

\eject
\figcaption{}
Three sets of the deflection angles $\alpha$ corresponding to three deflector masses
$10^{11}M_\sun$ (lower in red),
$10^{13}M_\sun$ (middle in green), and  $10^{15}M_\sun$ (top in blue) at redshift $z=1$
are shown as functions of the azimuthal angle $\tilde\phi_1$.
The thick lines are $|\alpha_{\rm total}|$ of Eq.\,(\ref{total}),
the short dashed lines are $|\alpha_{\rm static}|$ of Eq.\,(\ref{static}),
the dashed lines are $\alpha_{\rm expand}$ of Eq.\,(\ref{expand}),
and the  thin solid lines are the Einstein values $2r_s/r_0$.
All deflection angles are towards the deflector except
$\alpha_{\rm expand}$ which is away from the deflector, \ie in this plot
 $|\alpha_{\rm total}|=|\alpha_{\rm static}|-\alpha_{\rm expand}$.

\label{fig3}

\epsscale{.75}

\plotone{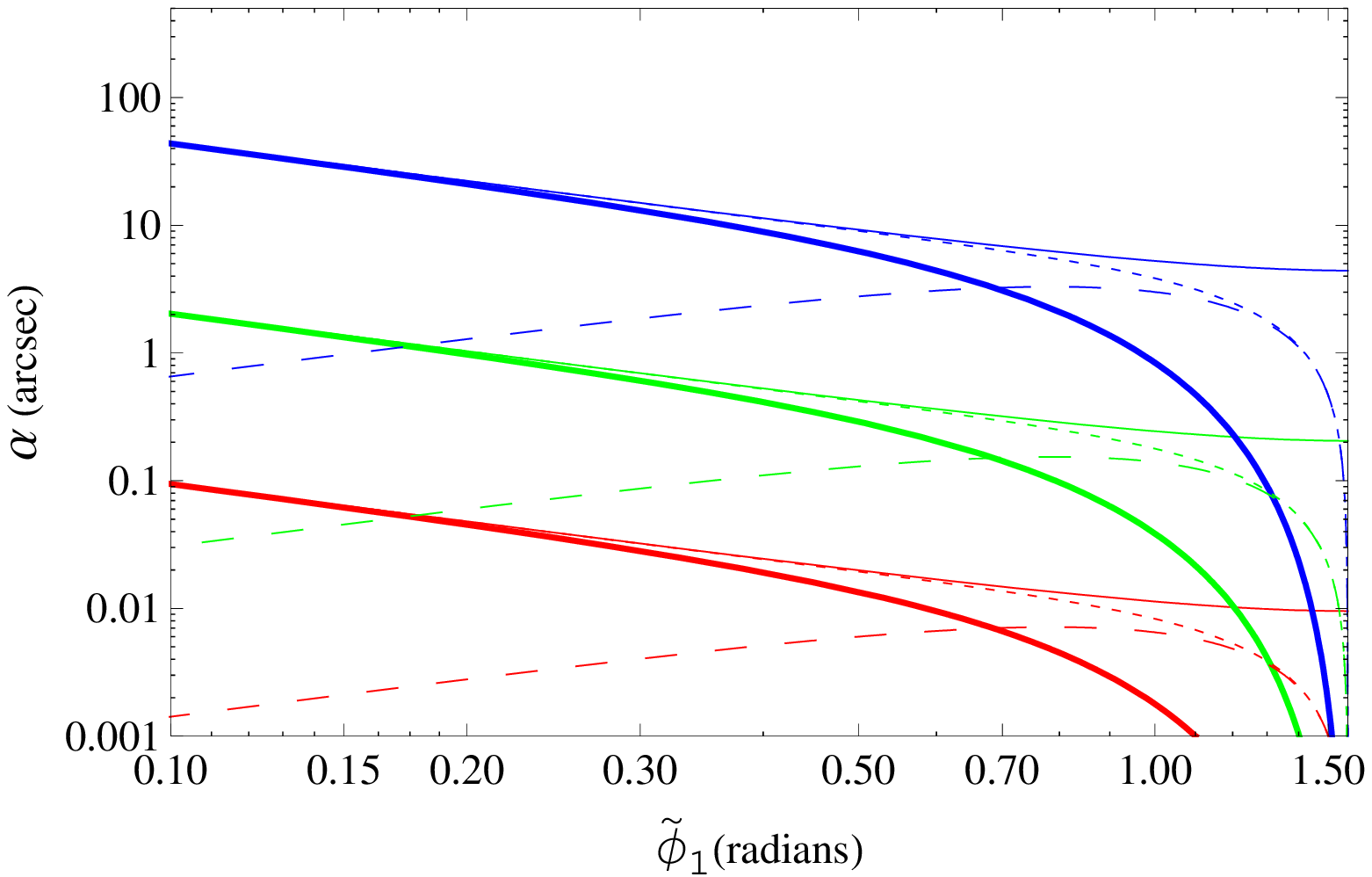}
\begin{center} Figure 3 \end{center}

\eject
\figcaption{}
The fractional difference of  the Einstein deflection angle
$2r_s/r_0$ and $\alpha_{\rm total}$ given in Eq.\,(\ref{total}),
\ie $(2r_s/r_0-|\alpha_{\rm total}|)/|\alpha_{\rm total}|$ as a function of $\tilde\phi_1$ in radians. For
the domain of  $\tilde\phi_1$ plotted and to the accuracy shown,
the fractional error is remarkably independent of the mass of the deflector ($10^{11}-10^{15}M_{\sun}$) and its redshift ($0<z<2$).

\label{fig4}

\epsscale{.75}

\plotone{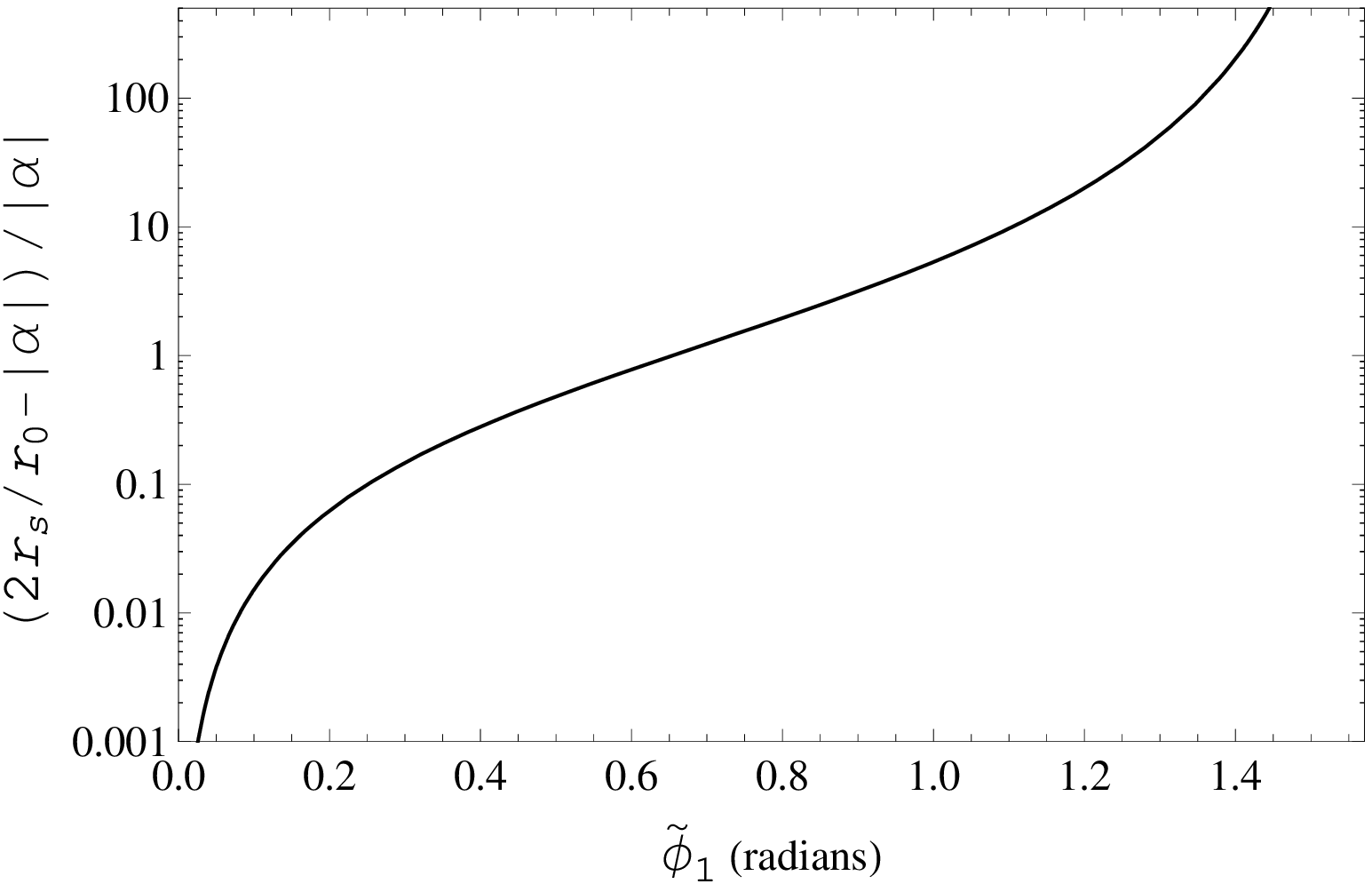}
\begin{center} Figure 4 \end{center}
\end{document}